\documentclass[twocolumn]{svjour3}         

\smartqed  

\usepackage{graphicx}
\usepackage{amssymb}
\usepackage{amsmath}
\usepackage{nicefrac}
\usepackage{color}
\usepackage{sansmath}
\usepackage{ulem}
\usepackage{comment}

\journalname{Tribology Letters}

\newcommand{\msout}[1]{\text{\sout{\ensuremath{#1}}}}

\begin{document}

\title{Percolation and Reynolds flow in elastic contacts of isotropic
and anisotropic, randomly rough surfaces}

\author{
 Anle Wang and Martin H. M\"user
}

\institute{
A. Wang \and 
M. H. M\"user \at
  {
    Department of Materials Science and Engineering,
    Saarland University, Campus C6 3,
    66123 Saarbr\"ucken, Germany. \\
    \email{martin.mueser@mx.uni-saarland.de}
  }
}

\date{Received: date / Accepted: date}

\maketitle

\begin{abstract}
In this work, we numerically study the elastic contact between isotropic and
anisotropic, rigid, randomly rough surfaces and linearly elastic counterfaces 
as well as the subsequent Reynolds flow through the gap between the two 
contacting solids.
We find the percolation threshold to depend on the fluid-flow direction when 
the Peklenik number indicates anisotropy unless the system size clearly 
exceeds the roll-off wave length parallel to the easy flow direction. 
A critical contact area near 
{0.415} is confirmed. 
Heuristically corrected effective-medium treatments satisfactorily provide
Reynolds fluid-flow conductances, e.g., 
for isotropic roughness, we identify accurate closed-form expressions,
which only depend on the mean gap and the relative contact area. 
\end{abstract}

\section{Introduction}

Predicting the leakage rate of seals requires the distribution of the 
interfacial separation between a surface and the seal to be known.
It can only be obtained reliably with accurate contact-mechanics models for the 
surface-seal system accounting for the microscopic roughness of solids.
The quantitative description of such systems can be said
to have had three births.
James Greenwood and J. B. Williamson (GW)~\cite{Greenwood1966} formulated the problem in 
1966 and suggested a solution to it in terms of non-interacting single-asperity contacts.
Bo Persson~\cite{Persson2001JCP} redefined the problem in 2001 by shifting the 
description and the solution of the contact problem from the real space to Fourier space, 
which ultimately lead to quite different results than those obtained by GW.
Meanwhile, Mark Robbins, who is honored in this issue of Tribology Letters,
lead the first efforts to rigorously model numerically
the multi-scale nature of roughness~\cite{Hyun2004PRE,Pei2005JMPS} and 
kept spearheading contact-mechanics simulations.
This gives us the chance to quickly sketch some of Mark's pioneering contributions to
contact mechanics. 

Mark understood much better than most of us that modeling is a two-step process:
``Reality'' is mapped onto mathematical equations in a first step, which then
need to be solved in a second step, typically by invoking additional approximations
to those while formulating the model.
He was one of the few who was strong in both and who would recognize that
scrutinizing what approximations can be made in each of the two steps is best
made separately.
For example, in his second work in the field of contact 
mechanics~\cite{Pei2005JMPS} he analyzed (i) to what
extent plastic deformation matters under what circumstances.
He identified rules for (ii) the range of validity of continuum theories for small-scale
contacts~\cite{Luan2005N} and 
worked out (iii) criteria for when randomly rough surfaces are (locally) 
sticky~\cite{Pastewka2014PNAS}.
In other work~\cite{Campana2008JPCM}, he found that (iv) stress and contact 
auto-correlation functions decay proportionally to
$\Delta r^{-(1+H)}$, as predicted by Persson~\cite{Persson2008JPCMb},
and not with $\Delta r^{-2\,(1+H)}$, as in bearing-area models like GW.
Mark also
(v) corroborated that Persson theory finds the correct load-displace-ment relation for 
randomly rough surfaces~\cite{Persson2007PRL}.
While it had already been established for moderate load when true contact
is spread across the interface~\cite{Almqvist2011JMPS,Dapp2012PRL},
Mark contributed to noticing that it also applies -- after some refinements -- 
when contact is localized near a single asperity~\cite{Pastewka2013PRE}.
%
%
%
%
The just-summarized insights that Mark contributed to the contact 
mechanics of nominally flat surfaces is but a small fraction of his overall 
contribution to tribology. 
%


The type of simulations that Mark conducted in his pioneering papers on 
nominally flat contacts has seen many subsequent works attempting to
pick up the crumbs that he left over, such as the subject of this study: 
contact-area percolation~\cite{Putignano2013TI,Yang2019PIME} in randomly-rough,
mechanical interfaces and the subsequent Reynolds flow through 
it~\cite{Dapp2012PRL,Dapp2016SR,PrezRfols2017PIME,PrezRfols2018PRSA,Vldescu2018JFE,PrezRfols2018L}.
This topic is merely one example for the use of full contact-mechanics simulations as 
starting points. 
%

The description of Reynolds flow in contacts between elastic, isotropic, randomly
rough surfaces appears to be well established, at least as long as the surface 
topographies obey the random-phase 
approximation~\cite{Dapp2012PRL,Persson2004JCP,Persson2008JPCMa,Lorenz2009EPL,Lorenz2010EPJE} 
but also for plastically deformed surfaces violating 
it~\cite{PrezRfols2018L,Persson2016TL}.
At small pressures, the fluid conductance disappears extremely quickly
with decreasing pressure until the dependence becomes roughly exponential
at moderate loads---as has been known experimentally for
a long time~\cite{Armand1964V}---before it disappears quickly on approach
to the percolation threshold~\cite{Dapp2016SR,PrezRfols2018L}. 
The exponential regime occurs for relative contact areas 
$a_\textrm{c}$ between a few percent up to close to the relative contact
area at the percolation threshold, $a_\textrm{c}^*$, which is believed to be
$0.42\pm 0.02$~ \cite{Dapp2012PRL,Dapp2016SR,PrezRfols2018L}.
While other values have also been proposed for $a_\textrm{c}^*$, it seems as if
the estimate $a_\textrm{c}^* \approx 0.4$ gets approached more closely as more care
is taken to simulate meaningful system sizes~\cite{Yang2019PIME}.
Just below $a_\textrm{c}^*$, the conductance disappears with a power law
in $a_\textrm{c} - a_\textrm{c}^*$~\cite{Dapp2016SR,PrezRfols2018L}, thereby 
reflecting the way how individual critical constriction 
close~\cite{PrezRfols2018L,Persson2008JPCMa,Dapp2015EPL}.

In contrast to many other percolation problems, for which ``susceptibilities''
are dominated on large scales near the percolation threshold~\cite{Stauffer1994book}, 
prefactors to leakage rates near $a_\textrm{c}^*$ are determined on the small
scale as they depend on how flow is impeded locally by a few last critical
constrictions~\cite{Persson2008JPCMa,Dapp2016SR,Persson2020EPJE}.
Results obtained experimentally or in large-scale simulations are reproduced 
quite accurately in terms of effective-medium 
approaches~\cite{Dapp2012PRL,Lorenz2010EPJE,Persson2012EPJE} 
going back to Bruggeman~\cite{Bruggeman1935AP}.
Good-quality predictions can also be made with the concept of critical 
constrictions~\cite{Persson2008JPCMa,Persson2020EPJE}, unless relative contact
areas are very small.
%

In recent works, Persson extended his contact mechanics theory as well as his
subsequent Bruggeman and critical-constriction approaches to 
anisotropic roughness~\cite{Persson2012EPJE,Persson2020EPJE}.
He pursues various approximations to calculate the conductance
tensor for anisotropic media, in particular he assumes that 
(a) the percolation threshold does not depend on the direction in
anisotropic surfaces and 
(b) quantitative measures for the height anisotropy and the 
conductance anisotropy are similar. 

The assumption of an isotropic percolation threshold could be seen as 
potentially problematic for the following reason: 
the height profile for a Peklenik number $\gamma>1$ results from assuming
isotropic random roughness on a rectangular $L/\sqrt{\gamma} \times 
\sqrt{\gamma} \, L$ 
domain, which is stretched by a factor of $\sqrt{\gamma}$ parallel to 
the $x$-axis and compressed by the same factor parallel to the $y$-axis.
In the original domain, both contact patches and fluid channels percolate 
more easily parallel to the shorter edge of the rectangle than to the longer 
one. 
After the stretching/compressing transformation, contact patches and fluid 
channels tend to be stripes for anisotropic domains and percolation should 
be eased in the direction of stripes. 
Thus, even if the flow channel topography could be obtained by the same
stretching/compression operation that can be used to generate an anisotropic 
height profile, 
probabilities to have open or closed 
channels right at the percolation threshold would be directionally dependent. 
Superficial contemplation of flow-channel geometries in small systems
easily reinforces the impression that the critical contact area must
be greater in the easy direction than in the compressed direction,
see, e.g., Fig.~\ref{fig:heightProfile}.
However, the two-dimensional anisotropic bond percolation 
model~\cite{Redner1979JPA,Masihi2006PRE} exhibits a crossover between one and 
two-dimensional critical behavior at large system sizes. 
It yet remains to be seen if elastic contacts obey similar principles.
Additional complications may arise due to the possibility that the anisotropy
of the contact area and thus of the gaps could be larger than that of the
original heights, as is the case for elliptical Hertzian 
indenters~\cite{Greenwood1997TI}.
Quantifying the just-described effects does not appear to be a trivial
task, which is why we resort to large-scale simulations in this work. 

The remainder of this paper is organized as follows:
Sect.~\ref{sec:model} presents the pursued models, methods, and some theoretical
concepts including some addenda to the Bruggeman treatment
for isotropic and unisotropic leakage. 
Sect.~\ref{sec:results} contains the results and their discussion, while
final conclusions are drawn in Sect.~\ref{sec:discussion}.


\section{Model, Methods, Theory}
\label{sec:model}

Model, methods, and theory are mostly similar to those used in 
Refs.~\cite{Dapp2012PRL,Dapp2016SR,Dapp2015EPL}.
The main difference in the model is that we now also consider anisotropic surfaces
and that the used contact-mechanics code was optimized in the meantime. 
%
%
%
In this section, we focus on these up-dates as well as on aspects that
might have remained unclear in previous works along with some 
additions or corrections to existing Bruggeman approaches to leakage. 

\subsection{Model}
We consider an originally flat, linearly elastic body with contact modulus
$E^*$ in contact with a rigid randomly rough indenter on a periodically
repeated domain.
The height spectrum of the latter obeys the random-phase approximation, i.e.,
$\tilde{h}(\mathbf{q}) = \sqrt{C(q_\textrm{P})}\exp(i\,2\pi\,u_\mathbf{q})$,
where $\tilde{h}(\mathbf{q})$ is the Fourier transform of the height profile,
$u_\mathbf{q}$ a linear independent random number drawn on $(0,1)$, 
$\mathbf{q}$ a wave vector and $q_\textrm{P}$ its effective magnitude 
\begin{equation}
q_\textrm{P} = 
\sqrt{\gamma \, q_x^2+q_y^2 / \gamma}
\end{equation}
Here, $\gamma$ denotes the so-called Peklenik 
number~\cite{Peklenik1967PIMEConf,Li2004TL}, whose squared 
logarithm is a measure for anisotropy.
If $\gamma > 1$, ``stretching'' occurs parallel to the $x$-axis, while
it is parallel to the $y$-axis if $\gamma < 1$.
Grooves show up parallel to the stretching direction remotely similar to a situation
in which a surface was polished or scratched in that direction.

As default for the height spectrum, a continuous transition between the 
so-called roll-off regime at small wave vectors and the self-affine scaling at
large wave vectors is used~\cite{Majumdar1990,Palasantzas1993PRB,Persson2014TL,Jacobs2017STMP}, specifically
\begin{equation}
C(q) \propto \frac{\Theta(q_\textrm{s}-q)}
{\sqrt{1+(q/q_\textrm{r})^2}^{\,1+H}},
\end{equation}
where $H$ is called the Hurst exponent, while
$\Theta(...)$ is the Heaviside step function, which is unity for positive
arguments and zero else. 
$q_\textrm{s}$ and $q_\textrm{r}$ are $2\pi$ over short wavelength cutoff
and rolloff wavelength, which are denoted by $\lambda_\textrm{s}$ 
and $\lambda_\textrm{r}$, respectively. 
As default values for the height spectrum, we use
$\varepsilon_\textrm{t} \equiv \lambda_\textrm{r}/L = 
1/\{4\,{\max(\sqrt{\gamma},1/\sqrt{\gamma})}\}$ and 
$\varepsilon_\textrm{f} \equiv \lambda_\textrm{s}/\lambda_\textrm{r} = 1/16.$
The discretization is always made small enough to ensure the continuum
limit to be closely approached. 
We chose such relatively small system sizes, as large anisotropy place
large demands on the computational resources. 
More importantly, we ensured that conclusions do not change when the
dimensionless numbers $\varepsilon_\textrm{t,s}$ are decreased. 

The linearly elastic body and the rigid substrate interact through a 
non-overlap constraint.
They are squeezed against each other with a constant pressure $p$.
Once the contact is formed, the interfacial separation is stored and used
for further analysis of the Reynolds flow, i.e., we neglect the mechanical
pressure exerted by the fluid flow on the contact mechanics.
This is certainly a reasonable approximation for leakage problems, all the more 
the neglected coupling provides only a minor perturbation to the flow factor 
associated with an individual constriction,
while leaving exponents unchanged that define the power laws with which
flow approaches zero with increasing load~\cite{Dapp2015EPL}. 

The gap topography described by the field $u(\mathbf{r})$ defines the local 
fluid conductivity through the equation
\begin{equation}
\sigma(\mathbf{r}) = \frac{\textcolor{black}{u^3_\textrm{g}(\mathbf{r})}}
{12\eta},
\end{equation}
where $\eta$ denotes the viscosity of the fluid and $u_\textrm{g}(\mathbf{r})$
is the local interfacial separation, or brief, gap. 
The such obtained conductivity is then used in Reynolds thin-film equation
\begin{equation}
\mathbf{j}(\mathbf{r}) = \sigma(\mathbf{r}) \nabla p_f(\mathbf{r}),
\end{equation}
$
\mathbf{j}(\mathbf{r})$ being the areal current density and 
$\nabla p_f(\mathbf{r})$ the in-plane fluid-pressure gradient. 
Conductances are evaluated parallel to the two principal axes of the simulation cell.
Periodic boundary conditions are assumed in the direction normal to the 
fluid pressure  gradient to reduce finite-size effects.

Please note that the term roll-off wavelength and the value of $\lambda_\textrm{r}$ 
both refer by default to that of the original, isotropic surface.
When adding the clause \textit{in the easy direction}, we mean
$\lambda_\textrm{r}\,\max(\sqrt{\gamma},1/\sqrt{\gamma})$. 
In addition, the stand-alone term \textit{pressure} refers to the 
mechanical pressure squezing the elastomer against the rigid substrate.
The fluid pressure has the added clause \textit{fluid}.

\subsection{Methods}

The elastic contact problem is solved using Green's function molecular dynamics 
(GFMD)~\cite{Campana2006PRB}, which is used in combination with the 
fast-inertial relaxation algorithm (FIRE)~\cite{Bitzek2006PRL}
as described elsewhere~\cite{Zhou2019PRB}.

The cluster analysis is based on the Hoshen-Kopelman (HK)
algorithm~\cite{Hoshen1976PRB}, which identifies connected contact or
non-contact (fluid) clusters. 
If two nearest neighbors are either both contact or both non-contact they
belong to the same cluster. 
A cluster is called percolating when it extends from one side of the
domain to the other. 
Finally, the Reynolds equations is solved as described in 
Ref.~\cite{Dapp2015EPL} using the \texttt{hypre} 
package~\cite{Falgout2006} and the conjugate-gradient minimizer supplied
with it. 

All simulations and analysis were conducted with house-written codes.

\subsection{Theory}

Different aspects of 
the contact-mechanics theory by Persson relevant to this study
have been described numerous times.
Particularly relevant to this study are those works describing how to use the 
Bruggeman effec-tive-medium 
approximation~\cite{Lorenz2010EPJE,Dapp2012PRL,Persson2012EPJE,Bruggeman1935AP} 
using the gap-distribution function $\Pr(u_\textrm{g})$.

\subsubsection{Bruggeman effective-medium approach}

The self-consistent equation needed to be solved in order to estimate the  
conductance $\sigma_0$ in the Bruggeman formalism reads~\cite{Persson2012EPJE}
\begin{equation}
\frac{1}{\sigma_0} = \int d\sigma \, \Pr(\sigma) \, \frac{D}{\sigma + \sigma_0\,(D-1)},
\label{eq:BruggemanIso}
\end{equation}
where $\sigma$ is the conductivity at a given point, 
$\Pr(\sigma)$ is its distribution function, 
and $D$ the (effective) spatial dimension.
In the original treatment, $D$ is taken as the true spatial dimension,
i.e., $D = 2$ for an interfacial leakage problem.

The conductance approaches zero when the probability for
zero conductivity exceeds $(D-1)/D$. 
This result inspired Dapp {\it et al.}~\cite{Dapp2012PRL} to use heuristically 
an effective spatial dimension
\begin{equation}
D = \frac{1}{1-a_\textrm{c}^*}
\end{equation}
in the Bruggeman effective-medium approach. 
In a similar spirit, 
Persson generalized Eq.~(\ref{eq:BruggemanIso}) to
\begin{equation}
\frac{1}{\sigma_{x,y}} = \int d\sigma  \, \Pr(\sigma) \,
\frac{D-1 + \tilde{\gamma}^{\pm 1} }
{\sigma + \sigma_{x,y} \,(D-1)\,\tilde{\gamma}^{\pm 1}}
\label{eq:BruggemanUniso}
\end{equation}
with
\begin{equation}
\tilde{\gamma} = \gamma \sqrt{\sigma_y/\sigma_x}
\end{equation}
for anisotropic media characterized by $\gamma \ne 1$. 
One flaw of  Eq.~(\ref{eq:BruggemanUniso}) is that it predicts 
different flows in $x$ and $y$ direction when the entire contact
is assigned the same microscopic conductivity when $D \ne 2$ is used, i.e.,
if $\Pr(\sigma) = \delta(\sigma-\sigma_0)$. 
To fix this, we modified Eq.~(\ref{eq:BruggemanUniso}) to
\begin{equation}
\frac{1}{\sigma_{x,y}} = \int d\sigma  \, \Pr(\sigma) \,
\frac{1 + (D-1)\,\tilde{\gamma}^{\pm 1} }
{\sigma + \sigma_{x,y} \,(D-1)\,\tilde{\gamma}^{\pm 1}}.
\label{eq:BruggemanUniso2}
\end{equation}
As another consequence of our correction, the ratio $\sigma_x/\sigma_y$ now
approaches $\gamma^2$ for $a_\textrm{c} \to a_\textrm{c}^*$ as is the case 
in the anisotropic Bruggeman solution using $D = 2$, as well as in the
critical-constriction approach.

\subsubsection{Addendum to the Bruggeman approach on isotropic media}

Persson theory allows the relative contact 
area and the average gap $\bar{u}_\textrm{g}$ to be estimated 
as a function of pressure~\cite{Persson2007PRL,Almqvist2011JMPS,Dapp2012PRL,Yang2008JPCM}, even for generalized elastomers
such as thin sheets or elastomers with gradient elasticity~\cite{Muser2020TL}
much more easily than the gap distribution function. 
The question arises if simple order-of-magnitude estimates for the fluid
conductance can be obtained using only $\bar{u}_\textrm{g}$ and
the relative contact area. 
To achieve that, we rewrite Eq.~(\ref{eq:BruggemanIso}) as
\begin{equation}
\label{eq:BruggemanIso1}
\frac{1}{\sigma_0} = 
\frac{D\,a_\textrm{r}}{ \msout{\sigma +}\;\; \sigma_0\,(D-1)} +
\frac{D\,(1-a_\textrm{r})}{\sigma_\textrm{nc} + \sigma_0\,(D-1)},
\end{equation}
where the characteristic non-contact conductivity
$\sigma_\textrm{nc}$ is defined through
\begin{equation}
\label{eq:sigmaNC}
\frac{1}{\sigma_\textrm{nc} + \sigma_0\,(D-1)} =  
\left\langle
\frac{1}{\sigma + \sigma_0\,(D-1)},
\right\rangle_\textrm{nc}
\end{equation}
whose calculation necessitates knowledge of $\sigma_0$. 
Here, 
$\langle... \rangle_\textrm{nc}$ indicates an average over non contact.

Keeping $\sigma_\textrm{nc}$ 
\textcolor{black}
{formally (although it still
needs to be determined later), Eq.~(\ref{eq:BruggemanIso1}) can be solved
for $\sigma_0$ to yield}
\begin{eqnarray}
\sigma_0 & = & 
\sigma_\textrm{nc}\, 
\left(1 - a_\textrm{c}/a_\textrm{c}^*\right),
\end{eqnarray}
which, after insertion into 
\textcolor{black}
{Eq.~(\ref{eq:sigmaNC})},
leads to the following 
self-consistent equation for $\sigma_\textrm{nc}$:
\begin{equation}
\frac{1}{\sigma_\textrm{nc}} = 
\lim_{\sigma_\textrm{min}\to 0^+} 
\int_{\sigma_\textrm{min}}^\infty
\!d\sigma\, \Pr(\sigma)\,
\frac{1+\Delta \tilde{a}}{\sigma + \sigma_\textrm{nc}\Delta\tilde{a}}
\label{eq:newSelfCons}
\end{equation}
with $\Delta \tilde{a} \equiv (a_\textrm{c}^*-a_\textrm{c})/(1-a_\textrm{c}^*)$.
%
\textcolor{black}
{Since $\sigma_\textrm{nc}$ cannot diverge but only be 
finite or approach zero as $a$ tends to $a_\textrm{c}^*$, 
$\sigma_0$ 
is predicted to disappear linearly or even faster with 
decreasing distance from the
percolation threshold.}
 
\textcolor{black}
{The power law, with which $\sigma$ disappears as
$a_\textrm{r}^*$ is approached, depends}
on the shape of the gap distribution function $\Pr(u)$, 
from which
the conductivity distribution function follows via
\textcolor{black}{
$\Pr(\sigma) \msout{\,d\sigma} = (u^2/4\eta)\,\Pr(u)\, \msout{du}$}.
%
This is best discussed by approximating the gap distribution function
at small $u$ (which is decisive for whether or not the relevant integrals converge)
with $\Pr(u) \propto u^\mu$.
For $\mu>0$, $\sigma_\textrm{nc}(a_\textrm{c}^*)$ is 
easily shown to remain positive no matter how closely the lower integration 
bound $\sigma_\textrm{min}$ approaches zero,
while a positive exponent $\mu$ leads to an 
algebraic disappearance $\sigma_\textrm{nc}(a_\textrm{c}^*)$ in 
$\sigma_\textrm{min}$ for $u \to 0^+$.
For $\mu = 0$, the disappearance is only logarithmic.

In the case of short-range adhesion, adhesive necks form with an infinite slope
of the gap at the contact line close, which effectively induces $\mu > 0$. 
A $\sigma_0 \propto \Delta a' \equiv 1 - a_\textrm{c}/a_\textrm{c}^*$ 
dependence follows, as observed in simulations using short-range 
adhesion~\cite{Dapp2016SR,Dapp2015EPL}.
A faster than linear power-law disappearance of $\sigma_0$ in $\Delta {a}'$ 
is predicted for repulsive contacts for which $\mu < 0$. 
This is again consistent with previous 
simulations~\cite{Dapp2015EPL,Dapp2016SR} finding 
$\sigma_0 \propto \Delta {a'}^\beta$ with $\beta = 69/20$. 
Finally  bearing-area models implicitly assume $\mu = 0$ so that
logarithmic corrections would apply to the $\sigma_0 \propto \Delta a'$
proportionality. 
Although the critical behavior was not analyzed in detail, this is again
consistent with the observation that the conductance disappears substantially
more slowly with increasing contact area for overlap models than for
true elastic contacts~\cite{Dapp2012PRL}.

To account heuristically for any observed $\sigma_0 \propto \Delta {a'}^\beta$
dependence, we propose to use
\begin{equation}
\sigma_\textrm{nc} = \frac{\bar{u}_\textrm{g}^3}{12\,\eta\,}
\,f(a_\textrm{c}),
\end{equation}
where $f(a_\textrm{c})$ is a  correction function, or, depending on 
context or viewpoint also a ``fudge-factor'' function, into which correct criticality
can be encoded by choosing it as
\begin{equation}
f(a_\textrm{c}) = f_0 \,
\left(1 - a_\textrm{c}/a_\textrm{c}^*\right)^{\beta-1},
\end{equation}
where $f_0$ should be of order unity. 
A summary of the expected conductance reads,
\begin{equation}
\sigma \approx \frac{\bar{u}_\textrm{g}^3}{12\,\eta} \times
\begin{cases}
\Delta {a'}^{69/20} & \textrm{regular contacts} \\
\Delta {a'}        & \textrm{short-range adhesion} \\
\frac{\Delta {a'}} {1-\ln \Delta {a'}} & 
\textrm{bearing models,} 
\end{cases}
\label{eq:approxBruggeman}
\end{equation}
in which the prefactor $f_0$ (and in the case of the bearing-model an 
additive constant) was selected such that $\sigma$ assumes
a value of a ${\bar{u}_\textrm{g}^3}/{12\,\eta}$ at zero contact area,
while the finite-contact-area correction factor makes the conductance
disappear with the correct power law as $a_\textrm{c}^*$ is approach,
as deduced from the scaling of $\Pr(u_\textrm{g})$ in the limit 
of $u_\textrm{g} \to 0^+$. 
%

\subsubsection{Critical-constriction approach}

The critical-constriction approach to the leakage rate of seals was 
introduced in Refs.~\cite{Persson2004JCP,Persson2008JPCMa} and 
extended to anisotropic roughness recently~\cite{Persson2020EPJE}. 
The theory is based on the idea that fluid flow at 
contact areas close to the percolation threshold
is impeded by a random distribution of narrow constrictions
through which the fluid has to be squeezed and that the
dominant part of the fluid pressure falls off at these constrictions.
In an  interface, the number of such constrictions per unit length in $x$ 
and $y$ direction scales as $L_x / \sqrt{\gamma}$ and $\sqrt{\gamma}\,L_y$, 
respectively. 
In a percolating channel, fluid flows through some narrow constrictions with
random directions.
Even in the case of anisotropy, the flow in a macroscopically large system
has to go occasionally through a constriction in which the flow
direction is perpendicular to the easy flow direction. 

For $\gamma = 1$ and $L_x = L_y$, there are as many critical constriction
in the $x$-direction as in the $y$-direction so that an equivalent circuit 
diagram of the fluid flow consists of a single critical constriction.
This allows one to focus on just a single characteristic constriction and 
the question how it impedes fluid flow as a function of the geometry of this 
constriction.
We refer to the original literature~\cite{Persson2004JCP,Persson2008JPCMa}
for how to estimate its geometry theoretically and thus its resistance
to fluid flow. 

\section{Results}
\label{sec:results}

\subsection{Preliminary considerations}
\label{sec:preliminary}
\begin{figure*}[hbtp]
\includegraphics[width=0.95\textwidth]{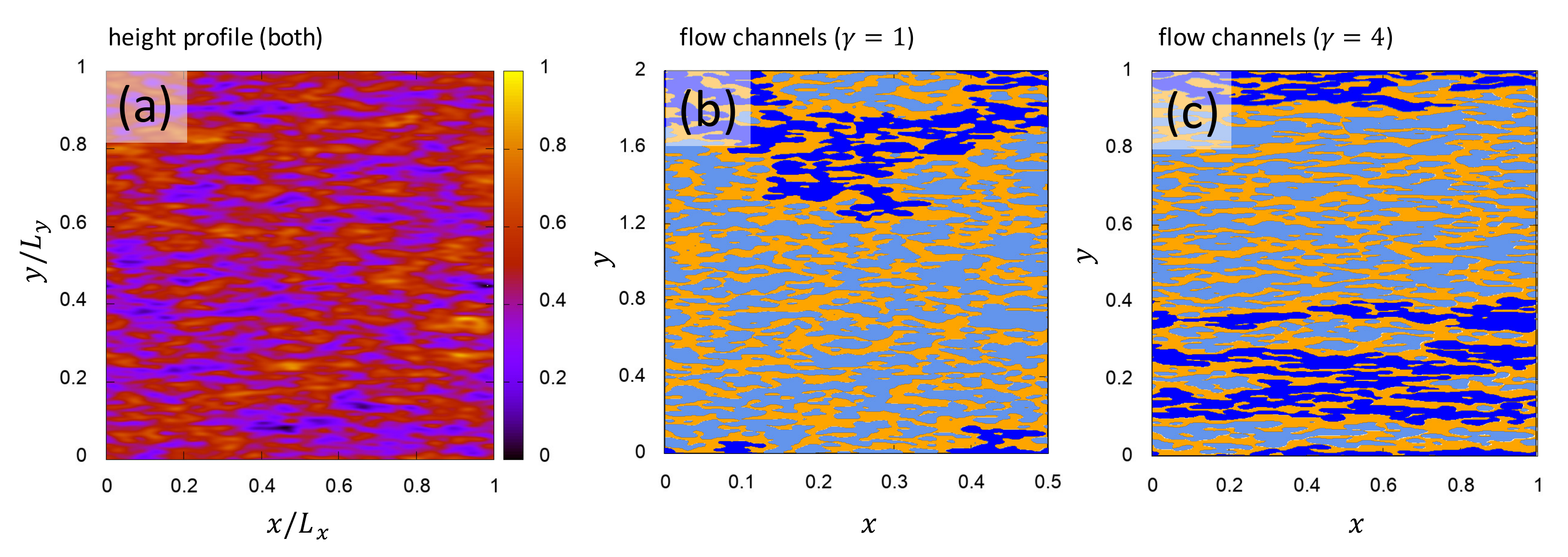}
\caption{\label{fig:heightProfile}
(a) Height profiles of an isotropic ($\gamma = 1$, $L_x = 0.5$, $L_y = 2$) and an
anisotropic ({$\gamma = 4$}, $L_x = L_y = 1$) unit-area surface, which is 
periodically repeated in the $y$ direction. 
Fluid channels (blue) and contact areas (orange) of the shown (b) isotropic and 
(c) anisotropic surface at $a_\textrm{c} \approx 0.4$.
Fluid channels percolating from left to right are rerpesented in dark blue,
other non-contact in light blue. 
Note that the $x$ and $y$ direction are not to scale for the isotropic surface,
which makes it appear anisotropic to the eye.
The height in panel (a) is stated in units of the maximum height. 
}
\end{figure*}

To set the stage for further discussion, flow channels for an isotropic but rectangular 
$(0.5\times 2)$ domain of a unit area are compared to those in a square, anisotropic 
domain, which is obtained from the former by scaling the $x$-direction with 
$\sqrt{\gamma} = 2$ 
and the $y$-direction with $1/\sqrt{\gamma}$.
This comparison is made in Fig.~\ref{fig:heightProfile}, which shows the height profile 
in panel (a) and contrasts the points of finite conductivity at a relative contact area 
of $a_\textrm{c} = 0.4$ for the isotropic and anisotropic surface in panels (b) and (c),
respectively. 
The original, rectangular domain is characterized by $H = 0.8$, $L_x = 0.5$, $L_y = 2$, 
$\lambda_\textrm{r} = 0.25$, and $\lambda_\textrm{s} = 0.025$. 
Both surfaces are at $39.8\pm 0.1$\% relative contact area. 

The expectation that stretching cannot change the percolation threshold, 
because the flow channel topology remains the same before and after the
stretching/ compression operation~\cite{Persson2020EPJE} is not fully supported
in the simulations. 
Although changes in the height profiles (not shown) are relatively minor, 
the fluid-channel topographies---and even topologies---shown in panels (b) and 
(c) of Fig.~\ref{fig:heightProfile} differ between the original and the 
stretched surfaces.
New percolating flow channels and percolating contact patches can open up after the 
stretching operation, while others disappear or merge. 
Both flow channels and contact patches of the elastic contact are even more
stretched than the height profile.
A related elongation of contact patches also occurs in isolated Hertzian contacts 
with elliptical indenters~\cite{Greenwood1997TI}.

A superficial contemplation of just this one random realization depicted in 
Fig.~\ref{fig:heightProfile} can easily convey the impression that an 
elastic contact characterized by the dimensionless numbers $H = 0.8$, 
{$\gamma = 4$}, 
and $a_\textrm{c} = 0.4$ should percolate parallel to the stretching direction
but not parallel to the orthogonal direction, even if the ratio of linear dimension
and $\lambda_\textrm{r}$ were larger than in the just-investigated example.
However, a numerical analysis and finite-size scaling ($\varepsilon_\textrm{t} \to 0$)
is required to test the validity of this expectation.

\subsection{Percolation threshold}

\begin{figure}[hbtp]
\vspace*{3mm}

\includegraphics[width=0.45\textwidth]{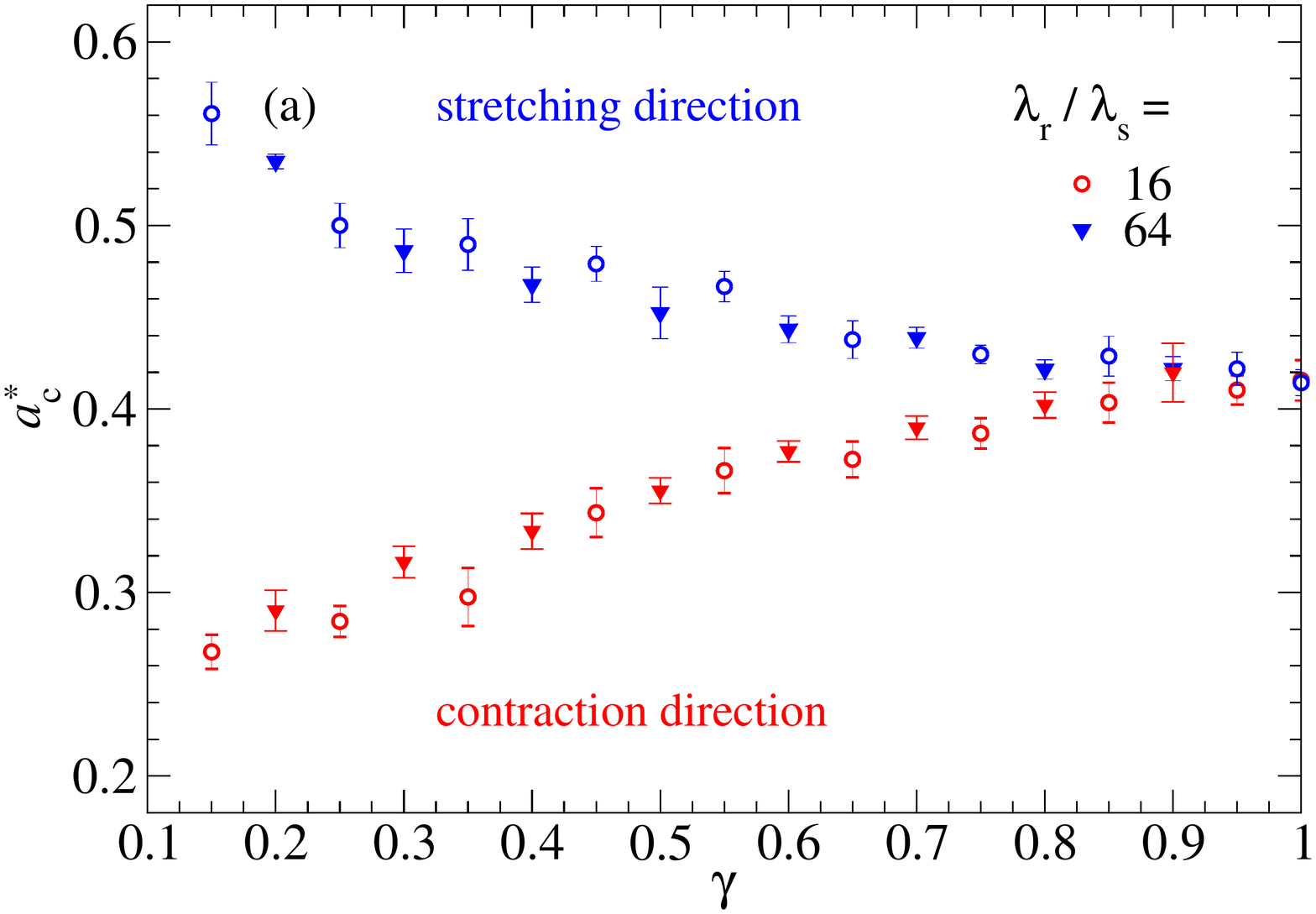}
\vspace*{3mm}

\includegraphics[width=0.45\textwidth]{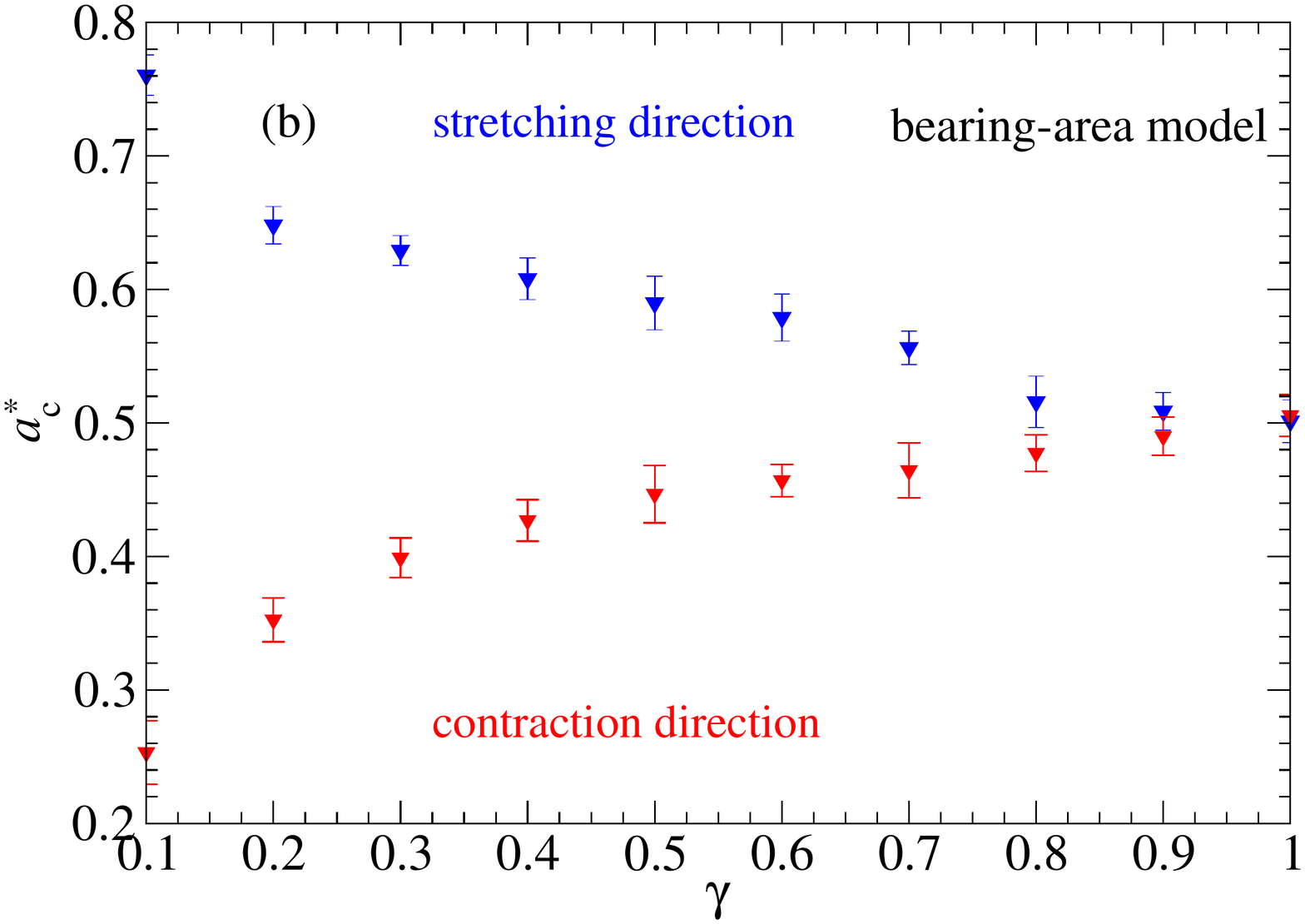}
\caption{ 
\label{fig:areaOfFractal}
Critical contact areas $a_\textrm{c}^*$ in the stretching (blue) and contraction
(red) directions as a function of the Peklenik number $\gamma$. 
(a) Elastic contact with two different ratios of 
$\varepsilon_\textrm{f} = \lambda_\textrm{s}/\lambda_\textrm{r}$ and
(b) bearing-area contact with $\varepsilon_\textrm{f} = 1/16$.
All cases correspond to $H = 0.8$ and 
$L/\lambda_\textrm{r} = 4/\sqrt{\gamma}$.
}
\end{figure}

In this section, we investigate how different dimensionless numbers
characterizing the surface topograhy affect the percolation threshold.
Towards this end, ten independent random realizations were 
typically set up to determine the order of magnitude of the stochastic 
error bars. 
Fig.~\ref{fig:areaOfFractal}(a) reveals that the percolation thresholds
$a_\textrm{r}^*$ along the two principal directions do not depend strongly
on the ratio
$\varepsilon_\mathrm{f} \equiv \lambda_\mathrm{s}/\lambda_\mathrm{r}$,
i.e., increasing it by a factor of 4 from 16 to 64 only has a relatively minor
effect,
\textcolor{black}
{which is clearly less than the stochastic error bar for
$\gamma$ close to unity.}
\sout{which remains within the stochastic error bars.}
%

An interesting feature revealed in Fig.~\ref{fig:areaOfFractal}(a) is that the
difference between the critical contact areas in the easy and the compression
direction increases with increasing anisotropy, although the system size
kept being increased proportionally to ${\max(1/\sqrt{\gamma},\sqrt{\gamma})}$. 
To test if this trend can be explained by the observation that the 
gap does not transform in the same self-affine fashion as the height,
we also computed $a^*_\textrm{r}$ along the two principal directions
for a bearing model, in which stretching and compressing is an affine
transformation.
In bearing-area models, contact is implicitly assumed to occur above a given 
substrate height and non-contact, i.e., open fluid flow channels, below it.
Fig.~\ref{fig:areaOfFractal}(b) reveals that the growth of asymmetry of
the critical contact areas with increasing $\gamma$ is similar 
for the bearing model as in the full elastic calculation. 
%
\textcolor{black}
{At moderate $\gamma$},
the main difference 
 between the two is a shift of 
$a_\textrm{c}^*(\gamma = 1) \approx 0.4$ in the elastic model to 
$a_\textrm{c}^*(\gamma = 1) = 0.5$ in the bearing model.

\begin{figure}[hbtp]
\includegraphics[width=0.45\textwidth]{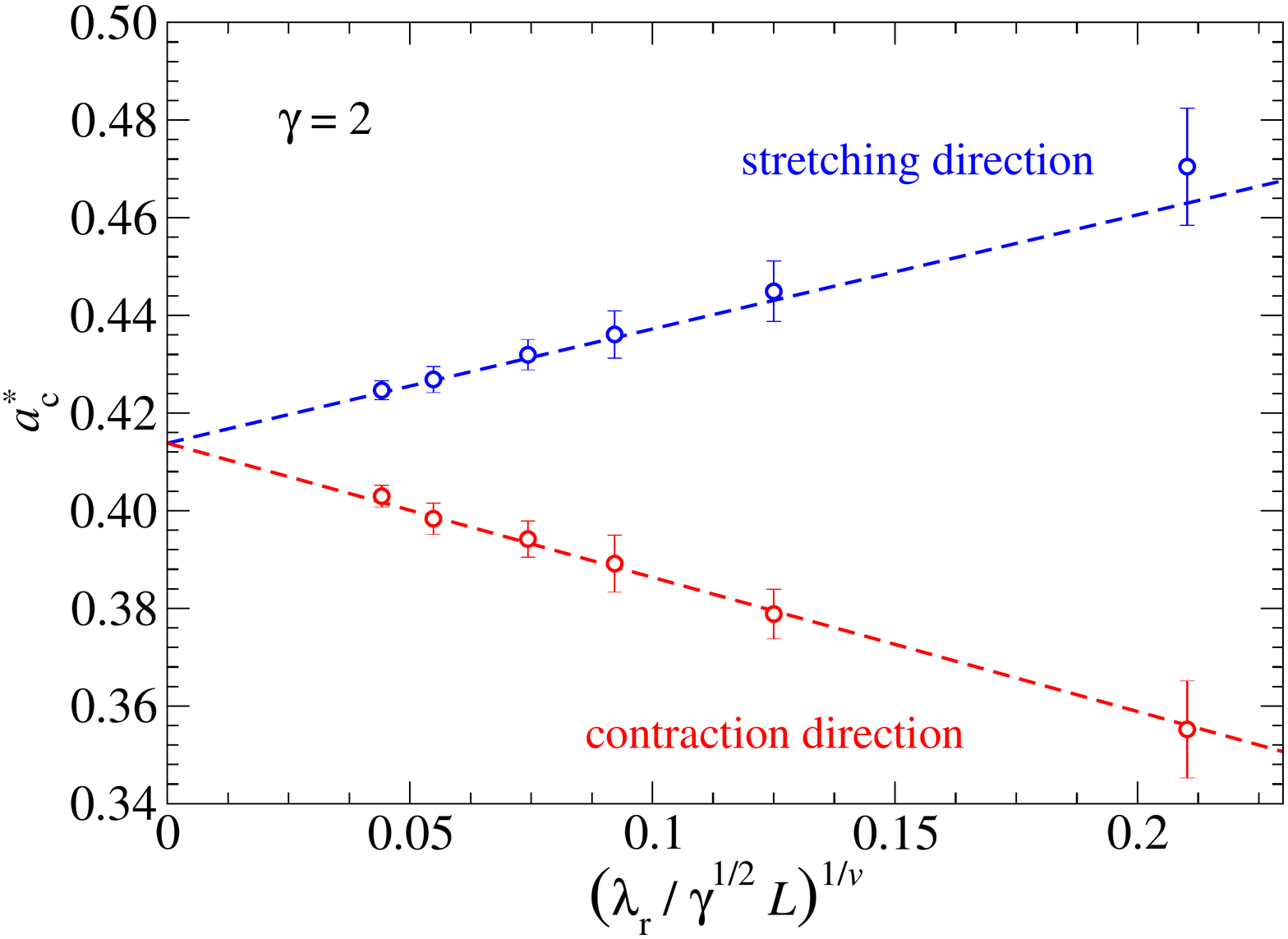}
\caption{ \label{fig:areaOfSize}
Size dependence of $a_\textrm{c}^*$ in the stretching (blue) and the
contraction (red) directions for a system with
a Hurst exponent $H = 0.8$. 
\textcolor{black}
{ Circles indicate constant $\gamma = 2$ and varying $L/\lambda_\textrm{r}$ ratios,
  while triangles assume a fixed ratio $L/\lambda_\textrm{r} = 32$  but
varying $\gamma$.
Black crosses show data for isotropic surfaces.
}
Lines are fits according to 
$a_\textrm{c}^*(\varepsilon_\textrm{c}) - a_\textrm{c}^* 
\propto \varepsilon^{1/\nu}$ with the random-bond-percolation-model 
exponent $\nu = 4/3$. 
}
\end{figure}

Although the system size was increased proportionally to the square root of the
(inverse) Peklenik number for the analysis presented in Fig.~\ref{fig:areaOfFractal},
the possibility remains that a further increase in system size suppresses
the observed anisotropy in $a_\textrm{c}^*$. 
This expectation is confirmed in Fig.~\ref{fig:areaOfSize}, which shows
that a unique percolation threshold of 
{$a_\textrm{c}^* 
\textcolor{black}{\approx 0.415 \pm 0.01}$}
is approached for the investigated
system with size corrections that are close-to-linear power laws in
$\varepsilon_\textrm{t}$.

The size scaling revealed in Fig.~\ref{fig:areaOfSize} is consistent with results 
for regular random-bond-percolation models. 
Its correlation length $\xi$ 
increases as  $\xi \propto 1/\vert a_\textrm{c} - a\vert^\nu$ and 
an exponent of $\nu = 4/3$ for an interfacial dimension of 
$D = 2$~\cite{Stauffer1979PR}.
Thus, the channels percolate along the easy direction at a finite size when 
$\xi \approx \sqrt{\gamma}\, L$ so that the size-dependent 
corrections of the relative contact area, $a^*_\textrm{c} - a^*$
satisfy
\begin{equation}
\sqrt{\gamma} L \propto \left\vert a^*_\textrm{c} - a^*_\textrm{c}(L)
\right\vert^{-\nu},
\end{equation}
which yields a size correction to $a^*_\textrm{c}$ of order
$L^{-1/\nu}$. 
Renormalization group theory arguments would then indicate
that the exponents describing size directions for the easy-flow direction 
and the contraction direction must be identical, however, 
the corrections must have opposite signs and may differ in magnitude.

It is currently not clear to us why the exponent $\nu$ are
identical 
\textcolor{black}
{or at least close} for the considered elastic contact 
problem and
the regular bond-percolation model, as there is no reason why percolation
in elastic contacts should be in the same universality class as 
random-bond percolation. 
In fact, the so-called Fisher exponent for the cluster-size distribution differs
between them.
It turns out $\tau = 187/92 \approx 2$ for regular bond 
percolation~\cite{Stauffer1979PR}
but $\tau \approx 2 - H/2$ for the contact-patch-size distribution
in repulsive, elastic contacts~\cite{Muser2018L}.

\begin{figure}[hbtp]
\vspace*{2mm}

\includegraphics[width=0.45\textwidth]{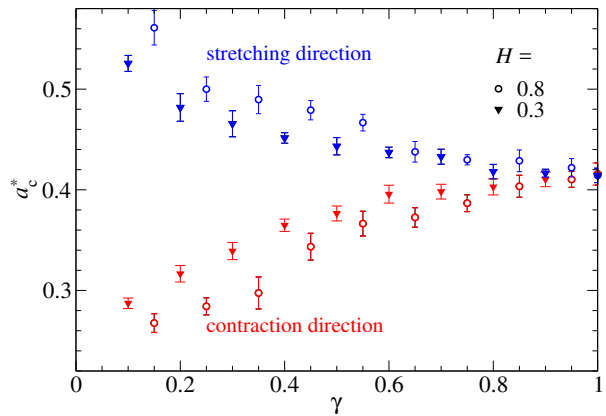}
\caption{ \label{fig:areaOfHurst}
Critical contact areas 
$a_\textrm{c}^*$ as a function of the Peklenik
number $\gamma$ for system sizes
$L = \lambda_\textrm{r}/\sqrt{\gamma}$ and $\varepsilon_\textrm{f} = 1/16$
for $H = 0.3$ (full triangles) and $H = 0.8$ (open circles). 
}
\end{figure}

\subsection{Reynolds flow}

\begin{figure}
\vspace*{2mm}

\includegraphics[width=0.45\textwidth]{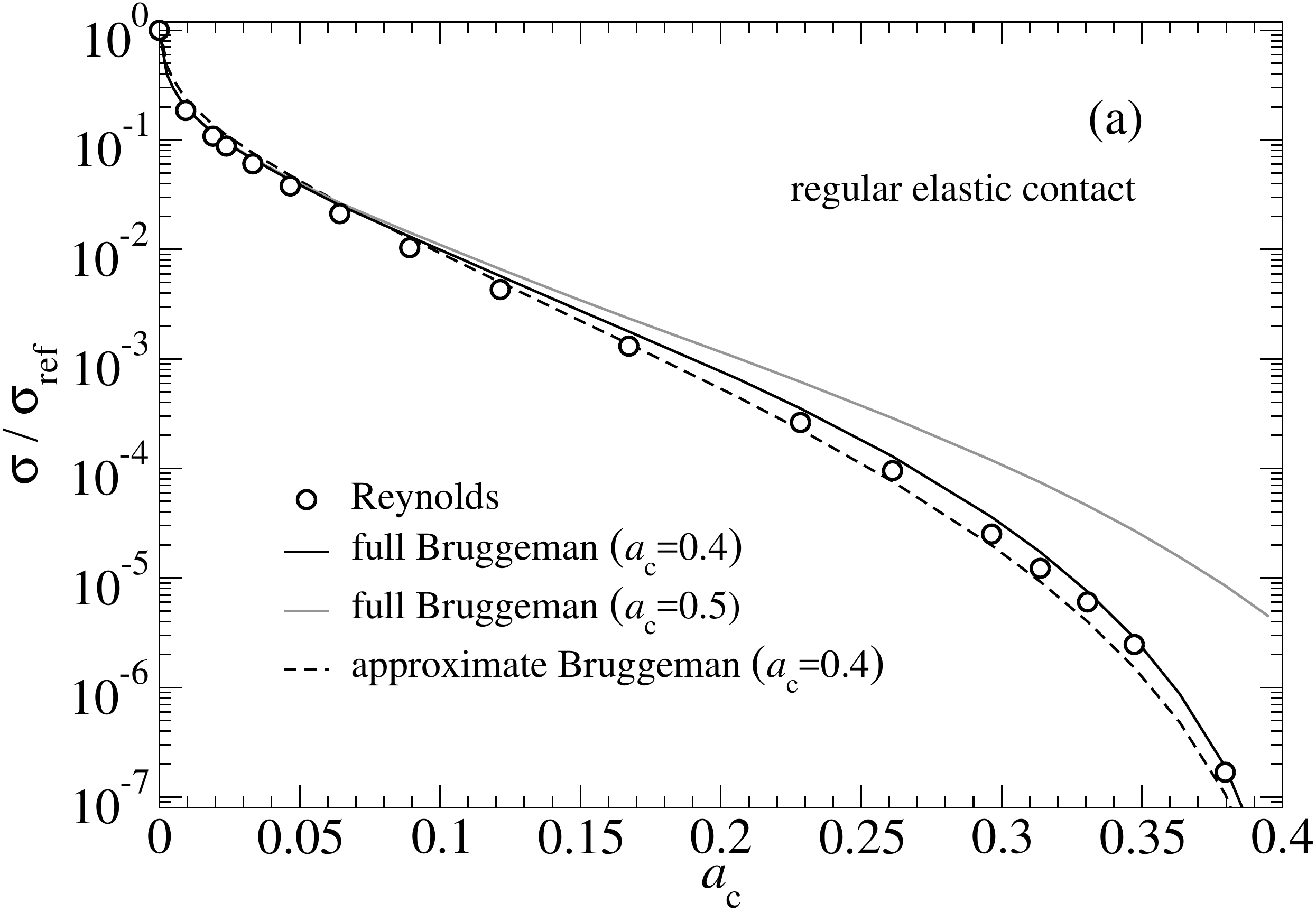}
\vspace*{2mm}

\includegraphics[width=0.45\textwidth]{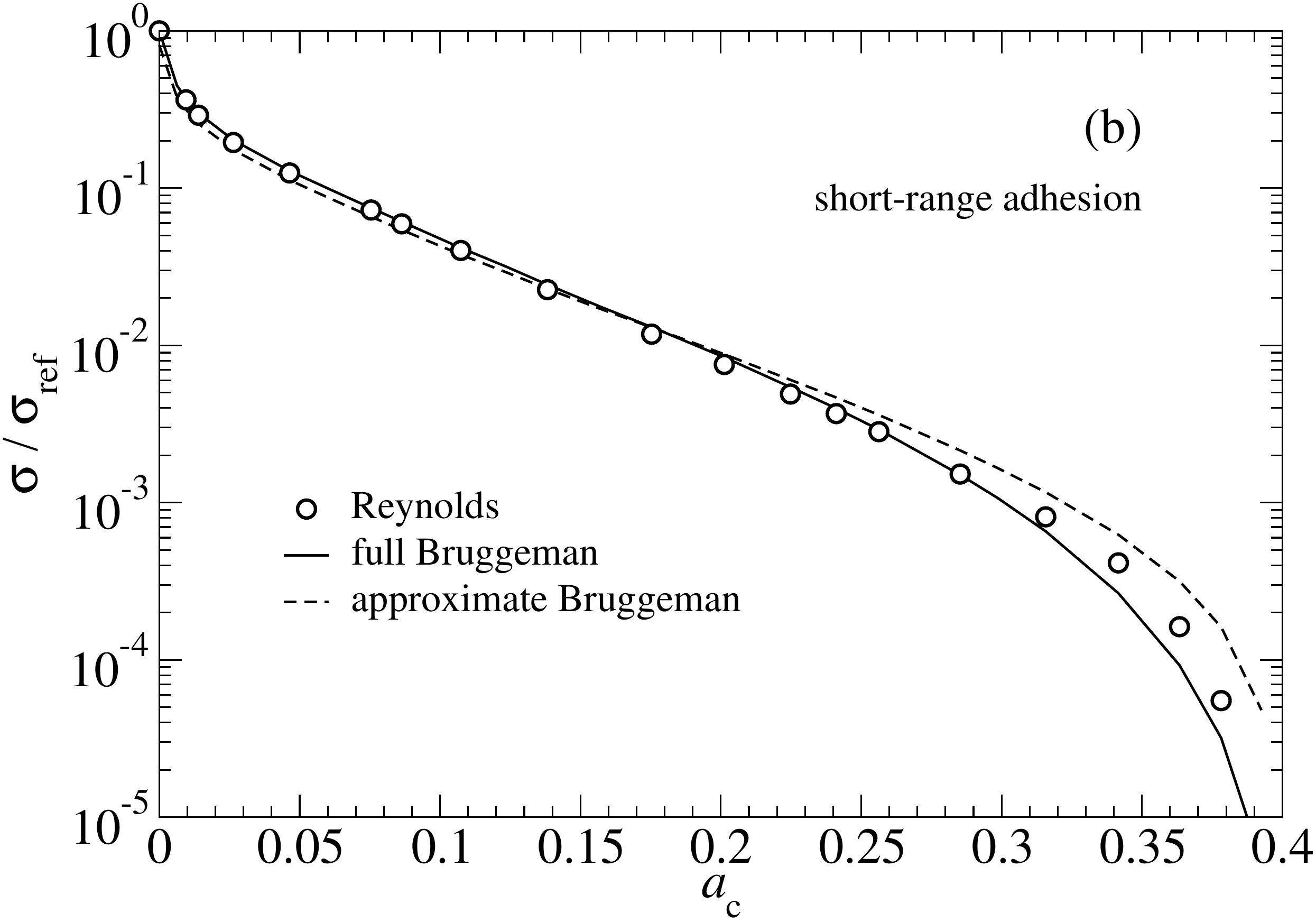}
\vspace*{2mm}

\includegraphics[width=0.45\textwidth]{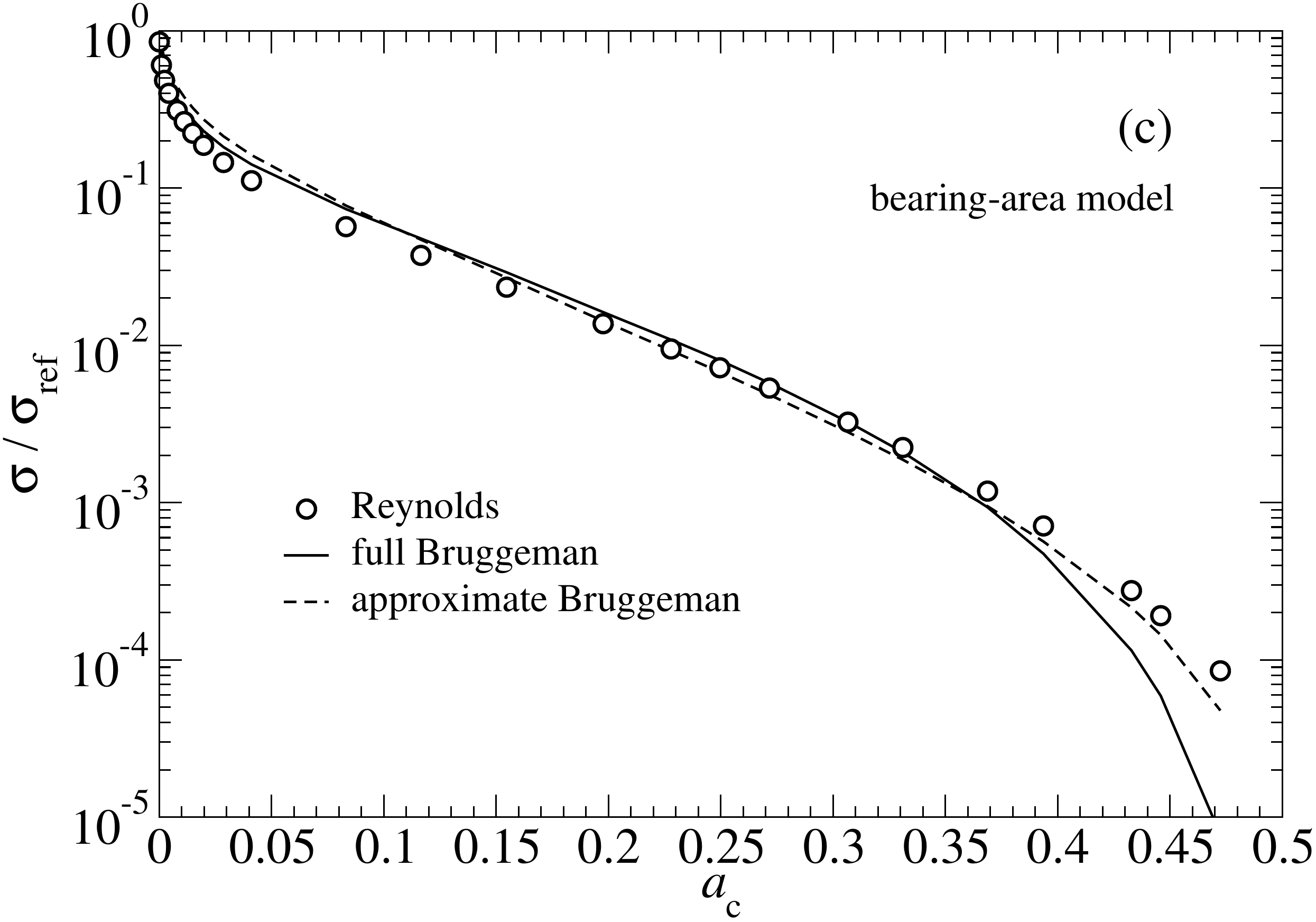}
\caption{\label{fig:flowIsotropic}
Fluid-flow conductance $\sigma$ as a function of
the relative contact area as obtained in full Reynolds calculation
(circles).  
Comparison is made to full Bruggeman effective-medium treatments
(full lines)
as well as to the approximations (dashed lines) proposed in 
Eq.~(\ref{eq:approxBruggeman})
for different contact models or treatments, i.e.,
(a) regular, elastic, non-adhesive contact,
(b) short-range adhesive contact, and
(c) bearing-area model.
%
%
In the case of (a), comparison is also made to a full Bruggeman
treatment assuming the critical contact area to be $a_\textrm{c}^* = 0.5$,
while the other lines in (a) and (b) are based on $a_\textrm{c}^* = 0.4$
The conductance is normalized to the same reference conductance
$\sigma_\textrm{ref}$, which is the one obtained in the full Reynolds
treatment at a relative contact area $a_\textrm{c} = 0^+$. 
}
\end{figure}

We start this section with the analysis of the Reynolds flow in isotropic
contacts.
It has already been demonstrated earlier~\cite{Dapp2012PRL,Dapp2016SR} that the 
Bruggeman effective-medium theory allows the ``exact'' Reynolds fluid conductance 
to be predicted quite accurately. 
In this paper, we test the validity of the closed-form analytical
expressions proposed for the isotropic conductance, which are summarized
in Eq.~(\ref{eq:approxBruggeman}).
In order to automatically yield good statistics, the system size was increased
from its default size to $L/\lambda_\textrm{r} = 16$, while the ratio
$\lambda_\textrm{r}/\lambda_\textrm{s} = 16$ was kept as before.
Fig.~\ref{fig:flowIsotropic} reveals that the analytical approximations
to the full Bruggeman theory are quite reasonable.
Relative deviations from either the numerically accurate solution of the full
Reynolds problem or the exact solution generally remain around 
{20\% in the shown domains,}
except for the adhesive case, where the full and the approximate Bruggeman
approach differed by a factor of two close to the percolation threshold.

The comparison between exact Reynolds and full as well as approximate
Bruggeman theory also adresses adhesive interface, which is 
shown in Fig.~\ref{fig:flowIsotropic}(b).
The strength and the range of adhesion were chosen such that it lead
to a non-negligible enhancement of local contact area, i.e., at zero load
we observed 1\% ``spontaneous'' relative contact area and to induce
a relative contact area of 10\% (40\%) only 1/8  (1/4) of the force was 
required as for its non-adhesive analogue. 
In more detail, the local Tabor parameter, as defined in 
Ref.~\cite{Muser2016TI}, was set to
{$\mu_\textrm{T} = 2$}, while the reduced surface energy, using the
so-called Pastewka-Robbins parameter, see Eq.~(16) in Ref.~\cite{Muser2016TI},
was {$\gamma_\textrm{PR} = 0.135$}.
Thus, no (local) stickiness can be expected despite the relatively large 
contact-area enhancement.
Also the ratio of surface energy $\gamma$ and the elastic energy per unit surface
needed to bring the two surfaces into the contact, $v_\textrm{ela}^\textrm{full}$, 
was well below unity, namely $\gamma/v_\textrm{ela}^\textrm{full} = 0.203$
further supporting the absence of hysteresis. 
In fact, there is a roughly constant, mere 10\% adhesion-induced reduction of 
the mean gap as a function of pressure in the studied range of forces but no signs 
of significant hysteresis. 
Thus, we would call the adhesion ``intermediate'', i.e., strong enough
to substantially increase the relative contact area but not so large as
to lead to substantial hysteresis. 

While the proposed dependence of conductance on mean gap and relative contact
area summarized in Eq.~(\ref{eq:approxBruggeman}) worked well for all
case studies performed for this study, it should be clear that
estimates can be rough close to the percolation threshold. 
This is because any short- but finite-range adhesion crosses over to
$\sigma \propto \Delta {a'}^{69/20}$ as the true percolation is approached,
see also Fig.~5 in Ref.~\cite{Dapp2016SR}. 
Likewise, if we had used very weak but zero-ranged adhesion, 
the trend might reverse, i.e., the conductance could be 
be proportional to 
$\Delta {a'}^{69/20}$ close but not too close to the 
percolation threshold but obey $\sigma \propto \Delta {a'}$
in the immediate vicinity of the percolation threshold. 
Thus, to be on the safe side, we recommend doing a full Bruggeman analysis 
(if possible), while its closed-form approximation can only 
provide
crude estimates for the conductance if the relative importance of adhesion 
is difficult to ascertain.

\begin{figure}[hbtp]
\includegraphics[width=0.45\textwidth]{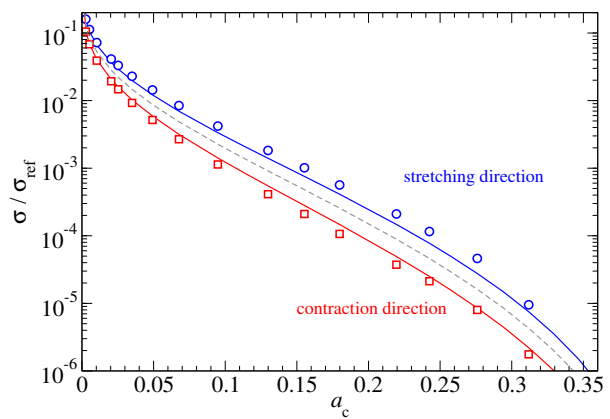}
\caption{\label{fig:anisoAcVSSigma}
Fluid conductance $\sigma$ in the stretching (easy) direction (blue)
and the contraction direction (red) for a system described by
a Peklenik number of $\gamma = 2$ 
\textcolor{black}
{and $L = 16\,\sqrt{\gamma}\,\lambda_\textrm{r}$}.
Results of a full Reynolds description are shown as symbols, while
predictions based on Eq.~(\ref{eq:BruggemanUniso2}) are shown in solid
lines.
The gray, dotted line is obtained using the Bruggeman approach for isotropic
surfaces.
}
\end{figure}

The last analysis of this work concerns the fluid conductance for
anisotropic surfaces for a system described by a Peklenik number of
$\gamma = 4$. 
Fig.~\ref{fig:anisoAcVSSigma} reveals that the generalization of the 
Bruggeman treatment for anisotropic elastic contacts conveys correct trends
but shows a somewhat weaker agreement with the full Reynolds calculations
than for isotropic surfaces. 

A quantitative analysis of the conductances reveals a  ratio of 
$\chi \equiv \sigma_x/\sigma_y = 8$, which is twice the 
\textcolor{black}
{theoretically expected number $\chi = \gamma^2$
using the height Peklenik number ($\gamma = 2$) to quantify 
the conductance anisotropy.}
%
A certain discrepancy from the theoretical expectation remains when
using instead the conductance Peklenik number of $\gamma_\sigma \approx 
\textcolor{black}
{2.5}$,
which we deduced from the direction-dependent conductivity auto-correlation
function (not shown). 
Thus using ``true'' conductivity Peklenik numbers leads to a predicted
ratio of 
\textcolor{black}
{$\chi \approx 6.25$, which reduces the error between
exact Reynolds calculations ($\chi \approx 8$) and effective
Bruggeman ($\chi = 4$) theory only by a little more than a factor of two.
}

\begin{figure}
\includegraphics[width=0.45\textwidth]{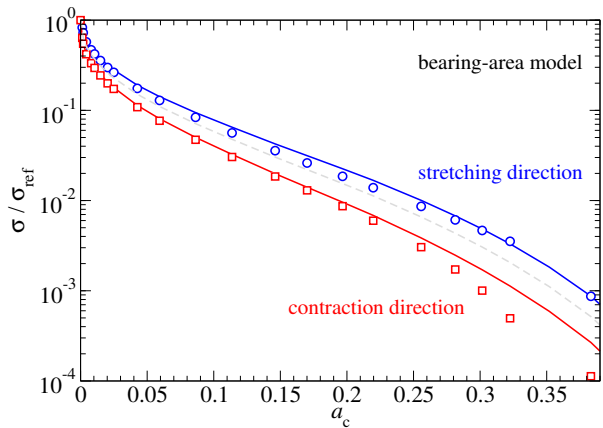}
\caption{ \label{fig:anisoAcVSSigmaBear}
Similar to Fig.~\ref{fig:anisoAcVSSigma} but for a bearing-area contact.
}
\end{figure}

To investigate the origin of the relatively large discrepancy  between
the exact Reynolds flow and the effec-tive-medium results for elastic
contacts, we also considered unisotropic bearing contacts, where conductivity
and height anisotropy are identical.
The results shown in Fig.~\ref{fig:anisoAcVSSigmaBear} reveal a similarly
close resemblance of the approximate solutions and the numerically exact 
results as for isotropic elastic contacts. 
We are tempted to explain the relatively poor performance of the
effective-medium theory near the percolation threshold with the following
argument:
the anisotropy that elastic deformation induces in addition to
the original stretching of the heights is particularly large
near the critical constrictions.

\section{Summary and Conclusions}
\label{sec:discussion}

In this work, we found that the relative contact area at which fluid
channels no longer percolate across a sufficiently large system
is $a_\textrm{c}^* = 
\textcolor{black}
{0.415 \pm 0.01}$ and that this value also holds
for surfaces with anisotropic random roughness.
This confirms Persson's conjecture that elastic anisotropic contacts have
a percolation threshold, which does not depend on the direction.
However, requirements on what is called ``sufficiently large'' are the
more stringent the greater the anisotropy. 
In addition, quantitative measures for anisotropy, such as the Peklenik
number, turn out larger for the fluid conductivity than for the height
of the randomly rough indenter,
at least within linearly elastic contact mechanics.
For bearing models, both yield identical Peklenik numbers. 

We also proposed a simplification as well as a minor correction to the 
Bruggeman effective-medium theory, which had been worked out by Persson
for the description of leakage in mechanical (elastic) contacts.
First, for isotropic contacts, we proposed quite simple, closed-form expressions
for the fluid-flow conductance in isotropic contacts, which necessitates only 
knowledge of the mean gap and the relative contact area as well as the type of 
contact (repulsive versus adhesive or in the odd case bearing-area contact) 
but it does not need as input the entire gap distribution function.
Second, we corrected the way in which an effective dimension is used in the
Bruggeman approach to anisotropic roughness in order to enforce the correct
percolation threshold. 
Both addenda to previous treatments were supported to our satisfaction by
full Reynolds simulations.

Finally, Persson's adaptation of the effective-medium theory to describe
direction-dependent conductances for anistropic media
works very well for bearing-area contacts, for which (a) the height-
and conductance Peklenik numbers are identical and (b) the percolation
threshold assumes the canonical value of $a_\textrm{c}^* = 1/2$. 
However, the generalization to anisotropic, elastic contacts is not quite
as satisfactory. 
It may well be that 
the way in which the correct percolation threshold is ``enforced'' for
elastic contacts through the use of an effective interfacial dimensions,
see Eq.~(\ref{eq:BruggemanUniso2}), can be further improved. 
Nonetheless, we find the approximate solution astoundingly good in all cases
given the simplicity of the effective-medium theory and 
the numerical complexity of a full Reynolds calculation.

\bibliographystyle{unsrt}

\end{document}